\documentclass[12pt]{article}
\usepackage{graphicx}
\usepackage{lscape}
\usepackage{citesort}
\usepackage{amssymb}
\usepackage{appendix}
\usepackage{multirow}
\usepackage{amssymb,epsfig}
\usepackage{psfrag}

\begin{document}
\def\qq{\langle \bar q q \rangle}
\def\uu{\langle \bar u u \rangle}
\def\dd{\langle \bar d d \rangle}
\def\sp{\langle \bar s s \rangle}
\def\GG{\langle g_s^2 G^2 \rangle}
\def\Tr{\mbox{Tr}}
\def\figt#1#2#3{
        \begin{figure}
        $\left. \right.$
        \vspace*{-2cm}
        \begin{center}
        \includegraphics[width=10cm]{#1}
        \end{center}
        \vspace*{-0.2cm}
        \caption{#3}
        \label{#2}
        \end{figure}
    }

\def\figb#1#2#3{
        \begin{figure}
        $\left. \right.$
        \vspace*{-1cm}
        \begin{center}
        \includegraphics[width=10cm]{#1}
        \end{center}
        \vspace*{-0.2cm}
        \caption{#3}
        \label{#2}
        \end{figure}
                }

\newcommand{\V}{{\cal V}}
\newcommand{\A}{{\cal A}}
\newcommand{\T}{{\cal T}}
\def\ds{\displaystyle}
\def\beq{\begin{equation}}
\def\eeq{\end{equation}}
\def\bea{\begin{eqnarray}}
\def\eea{\end{eqnarray}}
\def\beeq{\begin{eqnarray}}
\def\eeeq{\end{eqnarray}}
\def\ve{\vert}
\def\vel{\left|}
\def\ver{\right|}
\def\nnb{\nonumber}
\def\ga{\left(}
\def\dr{\right)}
\def\aga{\left\{}
\def\adr{\right\}}
\def\lla{\left<}
\def\rra{\right>}
\def\rar{\rightarrow}
\def\lrar{\leftrightarrow}
\def\nnb{\nonumber}
\def\la{\langle}
\def\ra{\rangle}
\def\ba{\begin{array}}
\def\ea{\end{array}}
\def\tr{\mbox{Tr}}
\def\ssp{{\Sigma^{*+}}}
\def\sso{{\Sigma^{*0}}}
\def\ssm{{\Sigma^{*-}}}
\def\xis0{{\Xi^{*0}}}
\def\xism{{\Xi^{*-}}}
\def\qs{\la \bar s s \ra}
\def\qu{\la \bar u u \ra}
\def\qd{\la \bar d d \ra}
\def\qq{\la \bar q q \ra}
\def\gGgG{\la g^2 G^2 \ra}
\def\q{\gamma_5 \not\!q}
\def\x{\gamma_5 \not\!x}
\def\g5{\gamma_5}
\def\sb{S_Q^{cf}}
\def\sd{S_d^{be}}
\def\su{S_u^{ad}}
\def\sbp{{S}_Q^{'cf}}
\def\sdp{{S}_d^{'be}}
\def\sup{{S}_u^{'ad}}
\def\ssp{{S}_s^{'??}}

\def\sig{\sigma_{\mu \nu} \gamma_5 p^\mu q^\nu}
\def\fo{f_0(\frac{s_0}{M^2})}
\def\ffi{f_1(\frac{s_0}{M^2})}
\def\fii{f_2(\frac{s_0}{M^2})}
\def\O{{\cal O}}
\def\sl{{\Sigma^0 \Lambda}}
\def\es{\!\!\! &=& \!\!\!}
\def\ap{\!\!\! &\approx& \!\!\!}
\def\md{\!\!\!\! &\mid& \!\!\!\!}
\def\ar{&+& \!\!\!}
\def\ek{&-& \!\!\!}
\def\kek{\!\!\!\!&-& \!\!\!\!}
\def\cp{&\times& \!\!\!}
\def\se{\!\!\! &\simeq& \!\!\!}
\def\eqv{&\equiv& \!\!\!}
\def\kpm{&\pm& \!\!\!}
\def\kmp{&\mp& \!\!\!}
\def\mcdot{\!\cdot\!}
\def\erar{&\rightarrow&}


\def\simlt{\stackrel{<}{{}_\sim}}
\def\simgt{\stackrel{>}{{}_\sim}}


\title{
         {\Large
                 {\bf
                     Study of the $\Lambda_{b} \to
                     N^\ast \ell^+ \ell^- $ decay in light cone sum rules
                 }
         }
      }

\author{\vspace{1cm}\\
{\small T. M. Aliev\,$^a\!\!$ \thanks {e-mail:
taliev@metu.edu.tr}\,\,,
T. Barakat\,$^b\!\!$ \thanks {e-mail:
tbarakat@ksu.edu.sa}\,\,, M. Savc{\i}\,$^a\!\!$ \thanks
{e-mail: savci@metu.edu.tr}} \\
{\small $^a$ Physics Department, Middle East Technical University,
06531 Ankara, Turkey} \\
{\small $^b$ Physics Department, King Saud University} \\
{\small Riyadh 11451, Saudi Arabia} }
\date{}

\begin{titlepage}
\maketitle
\thispagestyle{empty}

\begin{abstract}
\baselineskip 0.5 cm
The form factors of the $\Lambda_{b} \to N^\ast \ell^+ \ell^-$ decay are
calculated in the framework of the light cone QCD sum rules. In the
calculations the contribution of the negative parity $\Lambda_b^\ast$ baryon
is eliminated by constructing the sum rules for different Lorentz
structures. Furthermore the branching ratio of the semileptonic
$\Lambda_b \to N^\ast \ell^+ \ell^-$ decay is calculated. The numerical
study for the branching ratio of the $\Lambda_{b} \to N^\ast \ell^+ \ell^-$
decay indicates that it is quite large and could be measurable at future
planned experiments to be conducted at LHCb.
\\
\vspace{0.5cm} PACS numbers: 11.55.Hx; 13.30.Ce; 14.20.Gk\\
\\
\end{abstract}

\end{titlepage}

\section{Introduction}

Lately, exciting experimental results have been obtained in study of the rare
decays of the heavy $\Lambda_b$ baryon induced by the flavor changing
neutral currents. The rare $\Lambda_{b} \rar \Lambda \ell^+ \ell^-$  
decays induced by the $b \to s$ transition were observed by the CDF
\cite{Rhly01} and LHCb collaborations \cite{Rhly02}. Later the detailed
analyses of the differential branching ratio and various symmetries of the
$\Lambda_{b} \rar \Lambda \ell^+ \ell^-$ decay have been
performed performed at LHCb \cite{Rhly03}. The LHCb collaboration
firstly observed the rare $\Lambda_{b} \rar p \pi^- \mu^+ \mu^-$ decay
induced by the $b \to d$ transition \cite{Rhly04}. This observation
motivated the theoretical study of the $\Lambda_{b} \rar N \ell^+
\ell^-$ decay, induced also by the $b \to d$ transition. This decay   was
studied within the framework of the light cone QCD sum rules method 
(LCSR) in \cite{Rhly05}.
The light cone QCD sum rules method (LCSR) \cite{Rhly06,Rhly07} is hybrid of the
traditional SVZ sum rules \cite{Rhly08} and the methods used in hard
exclusive processes. The other interesting decays induced by the $b \to d$
transition are the $\Lambda_b \to$ nucleon resonance decays. The analysis of
these decays can provide complementary information about the
properties of the nucleon resonances in principle which could experimentally
be studied at LHCb. It should be noted here that the comprehensive study of
the nucleon resonance constitutes one of the main research directions of the
research program that is planned for future study at Jefferson Laboratory
\cite{Rhly09}.
The properties of the nucleon resonance $N^\ast$ in the
$\Lambda_{b(c)} \to N^\ast \ell \nu$ decay is investigated in framework
of the LCSR in \cite{Rhly10}.
The present work is devoted to the study of the rare $\Lambda_b \to N^\ast
\ell^+ \ell^-$ decay in the framework of the LCSR method. 

The paper is organized as follows: In section 2 the LCSR for the relevant
form factors appearing in the $\Lambda_b(\Lambda_b^\ast) \to N^\ast$
transitions are obtained. In section 3 present the numerical analysis of the
sum rules for the form factors. Using then the obtained results for the form
factors we estimate the decay widths of the   $\Lambda_b(\Lambda_b^\ast) \to
N^\ast \ell^+ \ell^-$ decays. This section ends with a conclusion.

\section{Form factors of the $\Lambda_b(\Lambda_b^\ast) \to \ell^+ \ell^-$
decay in LCSR}

In the present section we derive the LCSR for the transition form factors of
the  $\Lambda_b(\Lambda_b^\ast) \to \ell^+ \ell^-$ decay. Before giving the
details of the calculations few words about the notation should be
mentioned. In all further discussions the negative parity states of the
$\Lambda_b$ and $N$ baryons are denoted as $\Lambda_b^\ast$ and $N^\ast$,
respectively.
The $\Lambda_b(\Lambda_b^\ast) \to N^\ast \ell^+ \ell^-$ decay at the quark level
is described by the $b \to d$ transition. At the hadronic
level $\Lambda_b(\Lambda_b^\ast) \to N^\ast \ell^+ \ell^-$ decay is obtained by
sandwiching the transition current between the $\Lambda_b(\Lambda_b^\ast)$
and $N^\ast$ states. The corresponding form factors of the vector, axial
vector and tensor currents are defined as,
\bea
\label{ehly01}
\langle \Lambda_{Q}(p-q) \md  \bar b \gamma_\mu d \mid N^\ast(p)\rangle = \nnb \\
\bar {u}_\Lambda(p\kek q) \Big[f_{1}(q^{2})\gamma_{\mu}+{i} \frac{f_{2}(q^{2})}
{m_{\Lambda_b}}\sigma_{\mu\nu}q^{\nu} +
 \frac{f_{3}(q^{2})}{m_{\Lambda_b}} q^{\mu}\Big] \gamma_5 u_{N^\ast}(p)~, \\ \nnb \\
\label{ehly02}
\langle \Lambda_{Q}(p-q) \md  \bar b \gamma_\mu \gamma_5 d \mid N^\ast(p)\rangle = \nnb \\ 
\bar {u}_\Lambda(p\kek q) \Big[
g_{1}(q^{2})\gamma_{\mu}\gamma_5 + {i} \frac{g_{2}(q^{2})}{m_{\Lambda_b}}
\sigma_{\mu\nu} q^{\nu} \gamma_5
+ \frac{g_{3}(q^{2})}{m_{\Lambda_b}} q^{\mu}\gamma_5
\Big] \gamma_5 u_{N^\ast}(p)~, \\ \nnb \\
\label{ehly03}
\langle \Lambda(p-q)\md \bar b i \sigma_{\mu\nu}q^{\nu} (1+ \gamma_5)
d \mid N^\ast(p)\rangle = \nnb \\
\bar{u}_\Lambda(p\kek q)
\Big[\frac{f_{1}^{T}(q^{2})}{m_{\Lambda_b}}(\gamma_{\mu}q^2-q_{\mu}\not\!q)
+ {i}f_{2}^{T}(q^{2})\sigma_{\mu\nu}q^{\nu} \nnb \\
\ar  \frac{g_{1}^{T}(q^{2})}{m_{\Lambda_b}}(\gamma_{\mu}q^2-q_{\mu}\not\!q)\gamma_5
+{i}g_{2}^{T}(q^{2})\sigma_{\mu\nu} q^{\nu} \gamma_5 \Big]
\gamma_5u_{N^\ast}(p)~.
\eea
The form factors responsible for $\Lambda_b^\ast \to N^\ast$ transition can
be obtained from Eqs. (\ref{ehly01}), (\ref{ehly02}) and (\ref{ehly03}) with
the help of the following replacements:
$f_i \to \widetilde{f}_i$, $g_i \to \widetilde{g}_i$, $f_i^T \to
\widetilde{f}_i^T$, $g_i^T \to \widetilde{g}_i^T$, $m_{\Lambda_b} \to
m_{\Lambda_b^\ast}$, and $\bar {u}_\Lambda(p-q) \to \bar
{u}_{\Lambda^\ast}(p-q)\gamma_5$.

In order to derive the LCSR for these form factors we introduce the
correlation function      
\bea
\label{ehly04}
\Pi_\mu^j(p,q) =
i\int d^{4}xe^{iqx}\langle 0 \mid T\{\eta_{\Lambda_{b}}(0) J_{\mu}^j(x)\}\mid
N^\ast(p)\rangle~,
\eea
where $\eta_{\Lambda_{b}}$ is the interpolating current of the
$\Lambda_{b}(\Lambda_b^\ast)$ baryon, $J_{\mu}^j(x)$ is the heavy--light
transition current which is set to,
\bea
J_\mu^j = \left\{
  \begin{array}{ll}
\bar b \gamma_\mu(1-\gamma_5) d~,& j=I\\
\bar{b}~i\sigma_{\alpha\nu}q^{\nu} (1+ \gamma_5)d~, & j=II~. \nnb
\end{array}
\right.
\eea
In the calculations we use the following most general form of the 
interpolating current for the $\Lambda_{b}(\Lambda_b^\ast)$ baryon,
\bea
\label{ehly05}
\eta_{\Lambda_{b}}\es\frac{1}{\sqrt{6}}\epsilon_{abc}
\Bigg\{2\Big(\left[u^{aT}(x)Cd^{b}(x)\right]\gamma_{5}b^{c}(x)
+\beta\left[u^{aT}(x)C\gamma_{5}d^{b}(x)\right]b^{c}(x)\Big) \nnb \\
\ar \left[u^{aT}(x)Cb^{b}(x)\right]\gamma_{5}d^{c}(x) +
\beta\left[u^{aT}(x)C\gamma_{5}b^{b}(x)\right]d^{c}(x) \nnb \\
\ar \left[b^{aT}(x)Cd^{b}(x)\right]
\gamma_{5}u^{c}(x) +\beta\left[b^{aT}(x)C\gamma_{5}d^{b}(x)\right]u^{c}(x)\Bigg\}~,
\eea
where $a,~b$ and $c$ are the color indices, $C$ is the charge conjugation
operator, and $\beta$ is an arbitrary parameter with $\beta=-1$
corresponding to the Ioffe current.

The usual procedure in deriving the LCSR is to calculate the correlation
function given in Eq. (\ref{ehly04}) in two different domains. On one
side, insert a complete set of states with the quantum numbers of
$\Lambda_b$ between the two currents, and isolate the ground state
contribution. On the other side use the operator product expansion (OPE)
around the light cone where $(p=q)^2,~q^2 \ll 0$. These two representations
of the correlation function are then matched using the dispersion relations
and quark--gluon duality ansatz. Finally, applying the Borel transformation
in order to kill the the possible subtraction terms which could appear in the
dispersion relations, and to suppress the contributions from higher states.

It should be remarked here that the interpolating current $\eta_{\Lambda_b}$
has nonzero overlap not only with the $J^P=\ds {1\over 2}^+$ state but also
with the $J^P=\ds {1\over 2}^-$ state. It is shown in \cite{Rhly11} that the
mass difference between between the $J^P=\ds {1\over 2}^+$ and
$J^P=\ds {1\over 2}^-$ states is about $(200-300)~MeV$. For this reason the
contribution of the negative parity $\Lambda_b$ baryon  should properly be
taken into account.

After having mentioned these cautionary remarks we proceed to calculate the physical part of
the correlation function. Saturating Eq. (\ref{ehly04}) with the ground and first
excited $\Lambda_b$ baryon we get,
\bea
\label{ehly06}
\Pi_\mu^j(p,q)=\sum_i\frac{\langle 0\mid \eta_{\Lambda_{b}}(0)
\mid \Lambda_{b}^{i}(p-q,s)\rangle\langle
\Lambda_{b}^{i}(p-q,s)\mid  \bar{b} \Gamma_\mu^j d \mid
N^\ast(p)\rangle}{m_{i}^{2}-(p-q)^{2}}~,
\eea   
where
\bea
\Gamma_\mu^j = \left\{
  \begin{array}{ll}
\gamma_\mu (1-\gamma_5)~, & j=I~,\\
i\sigma_{\mu\nu} q^\nu(1+\gamma_5)~, & j=II~, \nnb
\end{array}
\right.
\eea
and summation is performed over the ground and first excited states of the
$\Lambda_b$ baryon.
The decay constants of the positive and negative parity $\Lambda_b$ baryons
are determined as,
\bea
\label{ehly07}
\langle 0 \mid  \eta_{\Lambda_b} \mid
\Lambda_b (p-q)\rangle \es \lambda_{\Lambda_b} u_{\Lambda_b}(p-q)~,\nnb \\
\langle 0 \mid  \eta_{\Lambda_b^\ast} \mid
\Lambda_b^\ast(p-q)\rangle \es \lambda_{\Lambda_b^\ast}\gamma_5
u_{\Lambda_b^\ast}(p-q)~.
\eea
Using the equation of motion
\bea
(\not\!{p} - m_{N^\ast}) u_{N^\ast}(p) = 0~,\nnb
\eea
and Eqs. (\ref{ehly01}--\ref{ehly03}) and (\ref{ehly07}), for the
phenomenological part of the correlation function we get,

\bea
\label{ehly08}
\Pi_{\mu}^{I}(p,q) \es \frac{\lambda_{\Lambda_b}}{m_{\Lambda_b}^{2}-(p-q)^{2}}\Bigg\{ 
f_{1}(q^{2})
\Big[(m_{\Lambda_b} + m_{N^\ast}) \gamma_\mu + \gamma_\mu \not\!q + 
 2 (p_\mu - q_\mu) I \Big]\gamma_5 \nnb \\
\ek \frac{f_{2}(q^{2})}{m_{\Lambda_b}}\Big[(m_{\Lambda_b} - m_{N^\ast})
[(m_{\Lambda_b} + m_{N^\ast}) \gamma_\mu + \gamma_\mu \not\!q]
 \gamma_5 + 2 \not\!q \gamma_5  p_\mu \nnb \\
\ar [(m_{\Lambda_b} - m_{N^\ast}) + \not\!q ]\gamma_5 q_\mu \Big]+
\frac{f_{3}(q^{2})}{m_{\Lambda_b}}
\Big[(m_{\Lambda_b} - m_{N^\ast}) - \not\!q\Big] \gamma_5 q_\mu \nnb \\
\ek g_{1}(q^{2}) \Big[
 (m_{\Lambda_b} - m_{N^\ast}) \gamma_\mu + \gamma_\mu \not\!q + 
 2 (p_\mu - q_\mu) I\Big] \nnb \\
\ar \frac{g_{2}(q^{2})}{m_{\Lambda_b}}
\Big[(2 p_\mu - q_\mu) \not\!q + 
 (m_{\Lambda_b} + m_{N^\ast}) [ (m_{\Lambda_b} - m_{N^\ast}) \gamma_\mu +
\gamma_\mu \not\!q - q_\mu I ]\Big] \nnb \\
\ar \frac{g_{3}(q^{2})}{m_{\Lambda_b}}
\Big[\not\!q - (m_{\Lambda_b} + m_{N^\ast}) \Big] q_\mu I \Bigg\} \nnb \\
\ar\frac{\lambda_{\Lambda_b^\ast}}{m_{\Lambda_b^\ast}^{2}-(p-q)^{2}}
\Bigg\{\widetilde{f}_1(q^{2}) \Big[(m_{\Lambda_b^\ast} - m_{N^\ast}) \gamma_\mu - 
\gamma_\mu \not\!q - 2 (p_\mu - q_\mu) I\Big]  \gamma_5 \nnb \\
\ar \frac{\widetilde{f}_2(q^{2})}{m_{\Lambda_b^\ast}} 	
\Big[(m_{\Lambda_b^\ast} + m_{N^\ast}) [(m_{\Lambda_b^\ast} - m_{N^\ast})
\gamma_\mu -  \gamma_\mu \not\!q ] \gamma_5  + 2 \not\!q p_\mu \nnb \\
\ar [(m_{\Lambda_b^\ast} + m_{N^\ast})  - \not\!q ] \gamma_5 q_\mu \Big]
+ \frac{\widetilde{f}_3(q^{2})}{m_{\Lambda_b^\ast}}
\Big[(m_{\Lambda_b^\ast} + m_{N^\ast}) I + \not\!q \Big] q_\mu \nnb \\
\ek \widetilde{g}_1(q^{2}) \Big[
 (m_{\Lambda_b^\ast} + m_{N^\ast}) \gamma_\mu - \gamma_\mu \not\!q -
 2 (p_\mu - q_\mu) I \Big] \nnb \\
\ek  \frac{\widetilde{g}_2(q^{2})}{m_{\Lambda_b^\ast}} \Big[
    (2 p_\mu - q_\mu) \not\!q  +  
    (m_{\Lambda_b^\ast} - m_{N^\ast}) [(m_{\Lambda_b^\ast} + m_{N^\ast})
 \gamma_\mu - \gamma_\mu \not\!q +  
       q_\mu I] \Big] \nnb \\
\ek  \frac{\widetilde{g}_3(q^{2})}{m_{\Lambda_b^\ast}}
\Big[\not\!q + (m_{\Lambda_b^\ast} - m_{N^\ast}) I\Big] q_\mu~,\\ \nnb \\
\label{ehly09}
\Pi_{\mu}^{II}(p,q) \es
\frac{\lambda_{\Lambda_b}}{m_{\Lambda_b}^{2}-(p-q)^{2}}\Bigg\{
{f_1^T(q^{2})\over m_{\Lambda_b}} 
\Big[ [(m_{\Lambda_b} + m_{N^\ast}) \gamma_\mu + \gamma_\mu\not\!q + 
 2 I p_\mu] \gamma_5 q^2 \nnb \\
\ar [(m_{\Lambda_b}^2  - m_{N^\ast}^2  - 
2 q^2)  - (m_{\Lambda_b} + m_{N^\ast}) \not\!q ] \Big] \gamma_5 q_\mu \nnb \\
\ek f_2^T(q^{2}) \Big[(m_{\Lambda_b} - m_{N^\ast}) [(m_{\Lambda_b} + m_{N^\ast}) \gamma_\mu + 
 \gamma_\mu \not\!q] \gamma_5 + 2 \not\!q \gamma_5 p_\mu \nnb \\
\ek [(m_{\Lambda_b} - m_{N^\ast}) + \not\!q] \gamma_5 q_\mu \Big] +
{g_1^T(q^{2})\over m_{\Lambda_b}}
\Big[ [(m_{\Lambda_b} - m_{N^\ast}) \gamma_\mu + \gamma_\mu \not\!q +
 2 I p_\mu] q^2 \nnb \\
\ek [(m_{\Lambda_b} - m_{N^\ast}) \not\!q - (m_{\Lambda_b}^2  - 
m_{N^\ast}^2 - 2 q^2)] q_\mu\Big] \nnb \\
\ek g_2^T(q^{2}) \Big[
(2 p_\mu - q_\mu) \not\!q  +  
 (m_{\Lambda_b} + m_{N^\ast}) [(m_{\Lambda_b} - m_{N^\ast}) \gamma_\mu + 
\gamma_\mu\not\!q - q_\mu] \Big] \Bigg\} \nnb \\
\ar \frac{\lambda_{\Lambda_b^\ast}}{m_{\Lambda_b^\ast}^{2}-(p-q)^{2}}\Bigg\{
{\widetilde{f}_1^T(q^{2})\over m_{\Lambda_b^\ast}}
\Big[[(m_{\Lambda_b^\ast} - m_{N^\ast}) \gamma_\mu - \gamma_\mu \not\!q  - 
2 I p_\mu] q^2 \gamma_5 \nnb \\
\ek [(m_{\Lambda_b^\ast}^2  - m_{N^\ast}^2  - 2 q^2) + 
(m_{\Lambda_b^\ast} - m_{N^\ast}) \not\!q ]\gamma_5 q_\mu\Big] \nnb \\
\ar \widetilde{f}_2^T(q^{2}) \Big[
(m_{\Lambda_b^\ast} + m_{\Lambda_b}^\ast) [(m_{\Lambda_b^\ast} - 
m_{\Lambda_b}^\ast) \gamma_\mu - \gamma_\mu\not\!q ] \gamma_5 + 
2 \not\!q \gamma_5 p_\mu \nnb \\
\ar [(m_{\Lambda_b^\ast} + m_{\Lambda_b}^\ast) - 
\not\!q]\gamma_5 q_\mu \Big] \nnb \\
\ar {\widetilde{g}_1^T(q^{2})\over m_{\Lambda_b^\ast}}\Big[
[(m_{\Lambda_b^\ast} + m_{N^\ast}) \gamma_\mu - \gamma_\mu \not\!q - 
2 I p_\mu] q^2 - [(m_{\Lambda_b^\ast} + m_{N^\ast}) \not\!q \nnb \\
\ar (m_{\Lambda_b^\ast}^2  - m_{N^\ast}^2  - 2 q^2)] q_\mu \Big] \nnb \\
\ar \widetilde{g}_2^T(q^{2}) \Big[
(2 p_\mu - q_\mu) \not\!q + 
(m_{\Lambda_b^\ast} - m_{N^\ast}) [(m_{\Lambda_b^\ast} + 
m_{N^\ast}) \gamma_\mu - \gamma_\mu \not\!q + I q_\mu] \Big]\Bigg\}~.
\eea

Now we turn our attention to the calculation of the correlation function
from the QCD side. At deep Eucledian domain $(p-q)^2,~q^2 \ll 0$ the product
of the two currents can be expanded around the light--cone $x^2 \simeq 0$.
After contracting the heavy quark fields which give the heavy quark
propagator, the matrix element
\bea
\epsilon^{abc} \langle 0 \mid u_\alpha^a (0) d_\beta^b (x) d_\lambda^c (0)
\mid N^\ast(p) \rangle~,\nnb
\eea
of the three quarks between the vacuum and the $N^\ast$ state is revealed.
Decomposition
of this matrix element in terms of the distribution amplitudes (DAs) with
increasing twist is given in \cite{Rhly12} (see Appendix A).

After contracting the heavy b--quark fields, the correlation takes the form,
\bea
\label{ehly10}
\Pi_{\mu}^j \es \frac{i}{\sqrt{6}} \epsilon^{abc}\int d^4x e^{iqx}
\Big\{\left[2 (C)_{\alpha\lambda} (\gamma_5)_{\rho\xi} +
(C)_{\alpha\xi} (\gamma_5)_{\rho\lambda} + (C)_{\xi\lambda} 
(\gamma_5)_{\alpha\rho}\right] \nnb \\
\ar \beta \left[2 (C \gamma_5)_{\alpha\lambda}(I)_{\rho\xi}
+ (C \gamma_5 )_{\alpha\xi}(I)_{\rho\lambda} + 
(C \gamma_5)_{\rho\lambda}(I)_{\alpha\rho} \right]\Big\}  \nnb \\
\cp \left(\Gamma^j\right)_{\sigma\beta}\left[S_b(-x)\right]_{\xi\sigma}
\langle 0 \mid  u_\alpha^a(0)
d_\beta^b(x) d_\lambda^c(0) \mid N^\ast (p)\rangle~.
\eea

\bea
\label{ehly11}
 S_b (x) \es  \frac{m_{b}^{2}}{4\pi^{2}}\frac{K_{1}(m_{b}\sqrt{-x^2})}
{\sqrt{-x^2}}-i \frac{m_{b}^{2}\not\!x}{4\pi^{2}x^2}K_{2}
(m_{b}\sqrt{-x^2})  \nnb \\
\ek i g_s \int \frac{d^4 k}{(2\pi)^4} e^{-ikx} \int_0^1 dv 
\Bigg[\frac{\not\!k + m_b}{( m_b^2-k^2)^2} G^{\alpha\beta}(vx)
\sigma_{\alpha\beta} + \frac{1}{m_b^2-k^2} v x_\alpha
G^{\alpha\beta} \gamma_\beta \Bigg]~,
\eea
where $K_i$ are the modified Bessel functions of the second kind, 
and $G_{\alpha\beta}$ is the gluon field strength tensor.

Using the expression of the heavy quark propagator and definition of the
matrix 
\bea
\epsilon^{abc} \langle 0 \mid u_\alpha^a (0) d_\beta^b (x)
d_\lambda^c (0) \mid N^\ast(p) \rangle~,\nnb
\eea
in terms of the DAs of the $N^\ast$ baryon, we can calculate the correlation
function from the QCD side. Note that, using the equation of motion
$(\not\!{p} - m_{N^\ast}) u_{N^\ast}(p) = 0$,
the correlation function can be
decomposed into six independent functions as follows, 
\bea
\label{ehly12}
\Pi_\mu^j(p,q) = \left[\Pi_1^j p_\mu + \Pi_2^j
p_\mu \not\!q + \Pi_3^j\gamma_\mu + \Pi_4^j\gamma_\mu
\not\!q + \Pi_5^jq_\mu + \Pi_6^jq_\mu\not\!q \right]
\gamma_5 u_{N^\ast}~.
\eea
After performing Borel transformation, these invariant functions in the
correlation function can in general be written as,
\bea
\label{ehly13}
\Pi_i^j\left[(p-q)^2, q^2\right] = \sum_{j=1,2,3} \int_0^1 { {\cal
D}_{ij} (x, q^2) \over \Delta^n}~,
\eea
where
\bea
\Delta = m_b^2 - (1-x) q^2 + x (1-x) m_{N^\ast}^2 - x (p-q)^2~.\nnb
\eea
The explicit forms of ${\cal D}_{ij}$ are quite lengthy and for this reason
we do not present them here. We can write (\ref{ehly13}) as a dispersion
integral in $(p-q)^2$ as follows,
\bea
\label{ehly14}
\Pi_i^j\left[(p-q)^2, q^2\right] = {1\over \pi} \int_{m_b^2}^\infty
{ds \over s-(p-q)^2}\mbox{Im}\Pi_i^j [(p-q)^2,q^2]~. 
\eea
Making the replacement $s(x) = m_b^2 - (1-x) q^2 + x (1-x) m_{N^\ast}^2 - x
(p-q)^2$, the denominator of Eq. (\ref{ehly13}) takes the form,
\bea
\Delta = x\left[s(x)-(p-q)^2\right]~. \nnb
\eea
Using the quark--hadron ansatz the contribution of the hadronic states can
be represented as,
\bea
\label{ehly15}
\int_{s_0}^\infty {ds\over s-(p-q)^2} \rho_i(q^2) \simeq
{1\over \pi} \int_{s_0}^\infty {ds\over s-(p-q)^2}
\mbox{Im}\Pi_i^j
(q^2)~,
\eea
where $s_0$ is the continuum contribution. Finally Borel
transformation can be performed on the hadronic and physical sides
with the help of the replacement
\bea
\label{ehly16}
{1\over s-(p-q)^2} \to e^{-s/M^2}~.
\eea
In implementing the Borel transformation and the continuum
subtraction we use the following relations,
\bea
\label{ehly17}
\int dx\frac{{\cal D}_{ij}^{(\alpha)}(x)}{\Delta}&\rar& \int_{x_{0}}^{1}\frac{dx}{x} {\cal D}_{ij}^{(\alpha)}(x)
e^{-s(x)/M^2}\nnb \\
\int dx\frac{{\cal D}_{ij}^{(\alpha)}(x)}{\Delta^2}&\rar& \frac{1}{M^2} \int_{x_{0}}^{1}
\frac{dx}{x^2}{\cal D}_{ij}^{(\alpha)}(x)e^{-s(x)/M^2}+ \frac{{\cal D}_{ij}^{(\alpha)}(x_{0})
e^{-s_0/M^2} }{m_{b}^2+x_{0}^{2}m_{N^\ast}^2-q^2}\nnb \\
\int dx\frac{{\cal D}_{ij}^{(\alpha)}(x)}{\Delta^3}&\rar& \frac{1}{2M^4} \int_{x_{0}}^{1}
\frac{dx}{x^3}{\cal D}_{ij}^{(\alpha)}(x)e^{-s(x)/M^2}+\frac{1}{2M^2} \frac{{\cal D}_{ij}^{(\alpha)}(x_{0})
e^{-s_0/M^2} }{x_{0}(m_{b}^2+x_{0}^{2}m_{N^\ast}^2-q^2)}\nnb \\
\ek\frac{1}{2} \frac{x_{0}^2e^{-s_0/M^2}}{m_{b}^2+x_{0}^{2}
m_{N^\ast}^2-q^2}\frac{d}{dx}\Big( \frac{{\cal D}_{ij}^{(\alpha)}(x)}{x(m_{b}^2+x^{2}m_{N^\ast}^2-q^2)}
\Big)\Big|_{x=x_{0}}~,
\eea
where $x_{0}$ is the solution of the equation
\bea
s_{0}=\frac{m_b^2-\bar{x}q^2+x\bar{x}m_{N^\ast}^2}{x}.\nnb
\eea
Equating the coefficients of the structures $p_{\mu}\gamma_5$,
$p_{\mu}\rlap/{q}\gamma_5$, $\gamma_{\mu}\gamma_5$,
$\gamma_{\mu}\rlap/{q}\gamma_5$, $q_\mu\gamma_5$, and
$q_{\mu}\rlap/{q}\gamma_5$ we get the following sum rules for the
invariant functions of the transition current ($\bar{b}\gamma_\mu d$),
\bea
\label{ehly18}
-2\lambda_{\Lambda_b}g_{1}(q^2)e^{-m_{\Lambda_b}^2/M^{2}}+2\lambda_{\Lambda_b^\ast}
\widetilde{g}_{1}(q^2)
e^{-m_{\Lambda_b^\ast}^2/M^{2}}\es\Pi_{1}^I(p,q)\nnb \\
2\lambda_{\Lambda_b}\frac{g_{2}(q^2)}{m_{\Lambda_b}}e^{-m_{\Lambda_b}^2/M^{2}}-
2\lambda_{\Lambda_b^\ast}\frac{\widetilde{g}_2(q^2)}{m_{\Lambda_b^\ast}}
e^{-m_{\Lambda_b^\ast}^2/M^{2}}\es \Pi_{2}^I(p,q) \nnb \\
-\lambda_{\Lambda_b}e^{-m_{\Lambda_b}^2/M^{2}}\Big\{(m_{\Lambda_b}-m_{N^\ast})\Big[g_{1}(q^2)
-\frac{g_{2}(q^2)}{m_{\Lambda_b}}
(m_{\Lambda_b}+m_{N^\ast})\Big]\Big\}-\nnb \\
\lambda_{\Lambda_b^\ast}e^{-m_{\Lambda_b^\ast}^2/M^{2}}\Big\{(m_{\Lambda_b^\ast}+m_{N^\ast})
\Big[\widetilde{g}_1(q^2)+
\frac{\widetilde{g}_2(q^2)}{m_{\Lambda_b^\ast}}(m_{\Lambda_b^\ast}-m_{N^\ast})\Big]\Big\}\es
\Pi_{3}^I(p,q)\nnb \\
- \lambda_{\Lambda_b}e^{-m_{\Lambda_b}^2/M^{2}}\Big[g_{1}(q^2)-
\frac{g_{2}(q^2)}{m_{\Lambda_b}}(m_{\Lambda_b}+
m_{N^\ast})\Big]+\nnb \\
\lambda_{\Lambda_b^\ast}e^{-m_{\Lambda_b^\ast}^2/M^{2}}\Big[\widetilde{g}_1(q^2)+
\frac{\widetilde{g}_2(q^2)}{m_{\Lambda_b^\ast}}(m_{\Lambda_b^\ast}-m_{N^\ast})\Big]\es
\Pi_{4}^I(p,q)\nnb \\
\lambda_{\Lambda_b}e^{-m_{\Lambda_b}^2/M^{2}}\Big[2 g_{1}(q^2)-
\frac{g_{2}(q^2)+ g_{3}(q^2)}{m_{\Lambda_b}}(m_{\Lambda_b^\ast}+m_{N^\ast})\Big]-\nnb \\
\lambda_{\Lambda_b^\ast}e^{-m_{\Lambda_b^\ast}^2/M^{2}}\Big[2\widetilde{g}_1(q^2)+
\frac{\widetilde{g}_2(q^2)+\widetilde{f}_3(q^2)}{m_{\Lambda_b^\ast}}
(m_{\Lambda_b^\ast}-m_{N^\ast})\Big]\es\Pi_{5}^I(p,q)\nnb \\
- \frac{\lambda_{\Lambda_b}}{m_{\Lambda_b}}e^{-m_{\Lambda_b}^2/M^{2}}
\Big[g_{2}(q^2)-g_{3}(q^2)\Big]+
\frac{\lambda_{\Lambda_b^\ast}}{m_{\Lambda_b^\ast}}e^{-m_{\Lambda_b^\ast}^2/M^{2}}
\Big[\widetilde{g}_2(q^2)-
\widetilde{g}_3(q^2)\Big]\es\Pi_{6}^I(p,q)~.
\eea
The results for the form factors induced by the $\bar{b}\gamma_\mu\gamma_5 d$
current can be obtained from 
Eq. (\ref{ehly18}) by making the following replacements:
$g_i\rar -f_i$, $\widetilde{g}_i\rar -\widetilde{f}_i$, 
$m_{N^\ast}\rar -m_{N^\ast}$, and $\Pi_i^I\rar \Pi_i^{I\prime}$.
 
The sum rules of the ($\bar{b}i\sigma_{\mu\nu} q^\nu d$) transition current can
be obtained in similar manner, which are given below,

\bea
\label{ehly19}
-2\lambda_{\Lambda_b}g_{2}^T(q^2)e^{-m_{\Lambda_b}^2/M^{2}}+2\lambda_{\Lambda_b^\ast}
\widetilde{g}_{2}^T(q^2)e^{-m_{\Lambda_b^\ast}^2/M^{2}}\es\Pi_{1}^{II}(p,q)\nnb \\
-\lambda_{\Lambda_b}e^{-m_{\Lambda_b}^2/M^{2}}\Big[\frac{g_{1}^T(q^2)}{m_{\Lambda_b}}(m_{\Lambda_b}-m_{N^\ast})-
g_{2}^T(q^2)\Big]-\nnb \\
\lambda_{\Lambda_b^\ast}e^{-m_{\Lambda_b^\ast}^2/M^{2}}\Big[\frac{\widetilde{g}_1^T(q^2)}{m_{\Lambda_b^\ast}}
(m_{\Lambda_b^\ast}+m_{N^\ast})+\widetilde{g}_2^T(q^2)\Big]\es\Pi_{2}^{II}(p,q)\nnb \\
\lambda_{\Lambda_b}e^{-m_{\Lambda_b}^2/M^{2}}\Big[\frac{g_{1}^T(q^2)}{m_{\Lambda_b}}
(m_{\Lambda_b}^2-m_{N^\ast}^2-2q^2)+g_{2}^T(q^2)(m_{\Lambda_b}+m_{N^\ast})\Big]-\nnb \\
\lambda_{\Lambda_b^\ast}e^{-m_{\Lambda_b^\ast}^2/M^{2}}\Big[\frac{\widetilde{g}_1^T(q^2)}{m_{\Lambda_b^\ast}}
(m_{\Lambda_b^\ast}^2-m_{N^\ast}^2-2q^2)-\widetilde{g}_2^T(q^2)(m_{\Lambda_b^\ast}-m_{N^\ast})\Big]\es\Pi_{3}^{II}(p,q)\nnb \\
\lambda_{\Lambda_b}e^{-m_{\Lambda_b}^2/M^{2}}(m_{\Lambda_b}-m_{N^\ast})\Big[\frac{g_{1}^T(q^2)}{m_{\Lambda_b}}q^2-
g_{2}^T(q^2)(m_{\Lambda_b}+m_{N^\ast})\Big]+\nnb \\
\lambda_{\Lambda_{b}^\ast}e^{\frac{-m_{\Lambda_{b}^\ast}^2}{M^{2}}}(m_{\Lambda_{b}^\ast}+m_{N^\ast})
\Big[\frac{\widetilde{g}_1^T(q^2)}{m_{\Lambda_b^\ast}}q^2+\widetilde{g}_2^T(q^2)(m_{\Lambda_b^\ast}-m_{N^\ast})
\Big]\es\Pi_{4}^{II}(p,q)~,
\eea
where $\Pi_{1}^{II}$, $\Pi_{2}^{II}$, $\Pi_{3}^{II}$, and $\Pi_{4}^{II}$
are the invariant functions for the structures $p_{\mu}\rlap/{q}\gamma_5$,
$q_{\mu}\rlap/{q}\gamma_5$, $q_\mu\gamma_5$, and $\gamma_{\mu}\gamma_5$, respectively.
The sum rules for the ($\bar{b}i\sigma_{\mu\nu}q^\nu\gamma_5 d $) transition current can be 
obtained from Eq. (\ref{ehly19}) by making the replacements $g_i^T\rar 
f_i^T$, $\widetilde{g}_i^T\rar \widetilde{f}_i^T$, $m_{N^\ast}\rar -m_{N^\ast}$, 
and $\Pi_i^{II}\rar \Pi_i^{II\prime}$. The explicit form of these
invariant functions are quite lengthy, and for this reason we do not present
them in this work.

Solving these equations we can eliminate the $\Lambda^\ast$ pole from the
sum rules. As the result we obtain the desired sum rules responsible for the  
$\Lambda_b \to N^\ast$ transition. In the next section we present our
numerical results on these form factors.

\section{Numerical analysis}

In this section we present our numerical results on the form factors that
describe the $\Lambda_b \to N^\ast$ transition. First let us
specify the input parameters which are needed in performing the numerical
calculations. The masses of the $\Lambda_b$ and $\Lambda_b^\ast$ baryons
which we use in our calculations are $\Lambda_b = 5.62~GeV$ and
$\Lambda_b^\ast= 5.85~GeV$, and the mass of the nucleon is $m_{N^\ast} =
1.52~GeV$ \cite{Rhly13}. The residues $\lambda_{\Lambda_b}$ and
$\lambda_{\Lambda_b^\ast}$ of the relevant baryons are taken from \cite{Rhly05}
having the values $\lambda_{\Lambda_b} = (6.5 \pm 1.5)\times 10^{-2}~GeV^3$
and $\lambda_{\Lambda_b^\ast} = (7.5 \pm 2.0)\times 10^{-2}~GeV^3$.
The mass of the $b$ quark is assigned to its $\overline{MS}$ given
as $m_b = (4.16 \pm 0.03)~GeV$ \cite{Rhly13}. The values of the quark condensates of
the light quarks are taken as, $\uu (1~GeV) = \dd (1~GeV) = -
(246_{-19}^{+28}~MeV)^3$. As has already been noted the main nonperturbative
parameters are DAs of the $N^\ast$ baryon. The expressions of the $N^\ast$
DAs, as well as the coefficients
$\phi_i^{(\pm,0)}$, $\psi_i^{(\pm,0)}$, and $\xi_i^{(\pm,0)}$, appearing in
the DAs are obtained in \cite{Rhly12} (also in
\cite{Rhly14,Rhly15,Rhly16,Rhly17,Rhly18}), and for completeness they are
presented in Appendix A.

The sum rules for the transition form factors contain three auxiliary
parameters: The Borel mass parameter $M^2$, the continuum threshold $s_0$ and
the arbitrary parameter $\beta$. The Borel mass parameter and the continuum  
threshold $s_0$ are determined from the criteria that the sum rule dictates,
i.e., the suppression of the contributions coming from the continuum states and
the higher  twist contributions should be satisfied. Our analysis shows that
the working regions of $M^2$ and $s_0$ lie in the region $M^2 = (10 \pm
5)~GeV^2$, $s_0 = (40 \pm 1)~GeV^2$, when aforementioned conditions are
fulfilled, and hence sum rules predictions are reliable. The final step
of our analysis is is the determination of the working region of the
parameter $\beta$. Our numerical study shows that when $-1 \le \cos\theta
\le -0.5$, where $\tan\theta = \beta$ the results for the residues and
masses are rather stable with respect to the variation of $\beta$, and we
choose $\beta = -1$.

The LCSR predictions are reliable up to the range $q^2 \le
q_{max}^2=(m_{\Lambda_b} - m_{N^\ast})^2$. In order to calculate the decay
width the LCSR predictions for the form factors need to be extrapolated to
the whole physical region. For this purpose we use the z--series
parametrization that is proposed in \cite{Rhly19},

\bea
\label{ehly21}
z(q^2, t_{0}) = \frac{\sqrt{t_{+}-q^2}-\sqrt{t_{+}-t_0}}{\sqrt{t_+-q^2}+
\sqrt{t_+-t_0}}~,
\eea
where $t_0 = (m_{\Lambda_b} - m_{N^\ast})^2$, $t_+ = (m_B + m_\pi)^2$.
The parametrization that best reproduces the form factors predicted by the
LCSR in the region $q^{2}\leq 11~ GeV^{2}$,
is given as
\begin{equation}
\label{ehly22}
f(q^2) = \frac{1}{1-q^2/(m_{\rm pole}^f)^2} \Big\{ a_0^f + a_1^f\:z(q^2,t_0) +
a_2^f \Big[z(q^2,t_0)\Big]^2 \Big\}~.
\end{equation} \nnb
For the pole masses we use,
\bea
 m_{pole} = \left\{
  \begin{array}{rl}
  m_{B^\ast} = 5.325~ GeV &  \!\mbox{for~~} f_1, f_2, f_{1}^T,
f_{2}^T;\widetilde{g}_1, \widetilde{g}_2, \widetilde{g}_{1}^T,
\widetilde{g}_{2}^T \\
    m_{B_1} = 5.723~ GeV & \!\mbox{for~~} g_1, g_2, g_{1}^T, 
g_{2}^T; \widetilde{f}_1, \widetilde{f}_2, \widetilde{f}_{1}^T, 
\widetilde{f}_{2}^T \\
     m_{B_0} = 5.749~ GeV & \!\mbox{for~~} f_3;~\widetilde{g}_3\\
    m_{B} = 5.280~ GeV  & \!\mbox{for~~} g_3;~\widetilde{f}_3 \nnb
  \end{array}
\right.
\eea
In Tables 1 and 2  we present the fit parameters $a_0$, $a_1$ and $a_2$ that
results from our numerical analysis.


\begin{table}[h]

\renewcommand{\arraystretch}{1.3}
\addtolength{\arraycolsep}{-0.5pt}
\small
$$
\begin{array}{|l|c|c|c|c|}
\hline \hline
          &         f_i(0)     &       a_0         &        a_1       &         a_2           \\ \hline
{f}_1     & -0.297  \pm 0.080  &  0.44 \pm 0.11    &  -0.17 \pm 0.03      &    4.16 \pm  0.40 \\
{f}_2     & -0.213  \pm 0.064  & -0.15 \pm 0.03    &  -1.36 \pm 0.26      &    5.22 \pm  0.90 \\
{f}_3     & -0.060  \pm 0.018  &  0.74 \pm 0.10    &  -7.24 \pm 1.20      &   16.22 \pm  2.20 \\
{g}_1     & -0.028  \pm 0.084  & -0.35 \pm 0.06    &   2.97 \pm 0.60      &   -6.87 \pm  0.95 \\
{g}_2     &  0.106  \pm 0.031  &  0.31 \pm 0.05    &  -0.93 \pm 0.16      &   -0.37 \pm  0.04 \\
{g}_3     & -0.017  \pm 0.005  & -0.19 \pm 0.04    &   1.66 \pm 0.31      &   -3.98 \pm  0.70 \\
{f}_1^T   & -0.0030 \pm 0.0008 &  6.29 \pm 1.40    & -62.42 \pm 6.30      &  154.71 \pm 18.00 \\
{f}_2^T   & -0.190  \pm 0.057  & -2.41 \pm 0.50    &  18.13 \pm 3.00      &  -35.77 \pm  7.10 \\
{g}_1^T   &  0.384  \pm 0.110  & -6.67 \pm 1.40    &  73.83 \pm 7.20      & -192.16 \pm 22.10 \\
{g}_2^T   & -0.190  \pm 0.056  &  1.34 \pm 0.30    & -13.40 \pm 2.10      &   29.03 \pm  6.00 \\
\hline \hline
\end{array}
$$
\caption{The parametrization of the form factors of the 
$\Lambda_b \to N^\ast \ell^+ \ell^-$ decay for LQSR--1}
\renewcommand{\arraystretch}{1}
\addtolength{\arraycolsep}{-1.0pt}
\end{table}


\begin{table}[h]

\renewcommand{\arraystretch}{1.3}
\addtolength{\arraycolsep}{-0.5pt}
\small
$$
\begin{array}{|l|c|c|c|c|}
\hline \hline
          &        f_i(0)      &        a_0       &         a_1     &        a_2      \\  \hline
{f}_1     & -0.202  \pm 0.060  & -0.27  \pm 0.06  &  -0.17 \pm 0.03 &   2.37 \pm 0.33 \\
{f}_2     & -0.0640 \pm 0.0018 & -0.060 \pm 0.012 &  -2.87 \pm 0.50 &   1.20 \pm 0.20 \\
{f}_3     &  0.0500 \pm 0.0015 &  0.60  \pm 0.14  &  -4.52 \pm 0.80 &   9.07 \pm 1.50 \\
{g}_1     & -0.144  \pm 0.043  & -0.38  \pm 0.06  &   1.65 \pm 0.22 &  -2.51 \pm 0.42 \\
{g}_2     &  0.062  \pm 0.002  &  0.24  \pm 0.04  &  -1.08 \pm 0.15 &  -1.17 \pm 0.20 \\
{g}_3     & -0.032  \pm 0.001  & -0.20  \pm 0.03  &   1.47 \pm 0.23 &  -3.11 \pm 0.50 \\
{f}_1^T   &  0.0210 \pm 0.0063 &  2.73  \pm 0.50  & -26.77 \pm 3.60 &  66.20 \pm 8.20 \\
{f}_2^T   & -0.176  \pm 0.052  & -1.12  \pm 0.22  &   7.66 \pm 1.80 & -14.89 \pm 2.30 \\
{g}_1^T   &  0.203  \pm 0.060  & -0.05  \pm 0.01  &   4.92 \pm 0.70 & -17.78 \pm 2.80 \\
{g}_2^T   & -0.176  \pm 0.052  &  0.030 \pm 0.006 &  -2.54 \pm 0.34 &   7.35 \pm 1.20 \\
\hline \hline
\end{array}
$$
\caption{The same as Table 1, but for LQSR--2}
\renewcommand{\arraystretch}{1}
\addtolength{\arraycolsep}{-1.0pt}
\end{table}

\clearpage

Using the definition  of the form factors the differential decay width is
calculated in the standard manner whose result is given below:
\bea
\frac{d\Gamma(s)}{ds} = \frac{G^2\alpha^2  m_{\Lambda_b}}{4096 \pi^5}|
V_{tb}V_{td}^*|^2 v \sqrt{\lambda(1,r,s)} \, \Bigg[ T_1(s) +
\frac{1}{3} T_2(s)\Bigg]~,
\eea
where $v_\ell=\sqrt{1- 4 m_\ell^2/q^2}$ is the lepton velocity,
$\lambda(1, r,  s)=1+r^2+s^2-2r-2 s-2r s$,
$s=q^2/m_{\Lambda_{b}}$, and  $r=m_{N^\ast}^2/m_{\Lambda_{b}}^2$,
$\alpha$ is the fine structural constant,
and the expressions of $\Gamma_1(s)$ and $\Gamma_2(s)$ are given in the
Appendix--B.

The dependence of the differential branching ratios on $q^2$ for the 
$\Lambda_b \to N^\ast \mu^+ \mu^-$ and 
$\Lambda_b \to N^\ast \tau^+ \tau^-$ decays, at $s_0=40~GeV^2$,
$M^2=25~GeV^2$ and $\beta=-1$ are presented in Figs. (1) and
(2), respectively. In these figures the graphical results predicted by the
LCSR--1 and LCSR--2 cases are shown together.

Performing integration over $s$ in the range $4 m_\ell^2 /m_{\Lambda_b}^2 
\le s \le (1-\sqrt{r})^2$, we obtain the branching
ratios for the $\Lambda_b \to N^\ast \ell^+ \ell^-~
(\ell=e,\mu,\tau)$ decays in the case when long distance effects are due
to the $J/\psi$ family, and these results are given in Table 3.

\begin{table}[h]

\renewcommand{\arraystretch}{1.3}
\addtolength{\arraycolsep}{-0.5pt}
\small
$$
\begin{array}{|l|c|c|}
\hline \hline
& \mbox{LCSR--1} & \mbox{LCSR--2} \\ \hline
\mbox{Br}(\Lambda_b \to N^\ast e^+ e^-)       &
(4.62 \pm 1.85) \times 10^{-8} &
(3.56 \pm 1.42)\times 10^{-8} \\
\mbox{Br}(\Lambda_b \to N^\ast \mu^+ \mu^-)   &
(4.25 \pm 1.50) \times 10^{-8} &
(3.25 \pm 1.24)\times 10^{-8} \\ 
\mbox{Br}(\Lambda_b \to N^\ast \tau^+ \tau^-) & 
(0.25 \pm 0.09 )\times10^{-8}  & 
(0.180 \pm 0.067)\times10^{-8}  \\
\hline \hline
\end{array}
$$
\caption{Branching ratios for the $\Lambda_b \to N^\ast \ell^+ \ell^-,~           
(\ell=e,\mu,\tau)$ decays}
\renewcommand{\arraystretch}{1}
\addtolength{\arraycolsep}{-1.0pt}
\end{table}

It follows from these results that, especially for the $e$ and $\mu$
channels, the branching ratios are quite large and could potentially
be measurable at
LHCb. The discovery of these decays would provide useful
information about the inner structure of the $N^\ast$ baryon.

Finally, a comparison of the
$\Lambda_b \to N \ell^+ \ell^-$ and 
$\Lambda_b \to N^\ast \ell^+ \ell^-$ decays shows that 
the central values of the branching ratio
of the former is approximately two (eight) times larger than the $\Lambda_b \to N^\ast e^+
e^-$ ($\Lambda_b \to N^\ast \tau^+ \tau^-$) decays. 

\section*{Conclusion}

In this work the transition form factors of the $\Lambda_b \to N^\ast \ell^+
\ell^-$ decay are estimated, which is an alternative approach to extract
information about the inner properties of the $N^\ast$ baryon. The
contribution coming from the $\Lambda_b^\ast$ baryon is 
eliminated by constructing the sum rules with the choice of several
different Lorentz structures. Using this result, we also calculate the
branching ratio of the $\Lambda_b \to N \ell^+ \ell^-$ and decays. We see
that the branching ratios of the $\Lambda_b \to N^\ast e^+ e^-$ and
$\Lambda_b \to N^\ast \mu^+ \mu^-$ seem to be large enough to be detected at LHCb.

\clearpage

\section{Acknowledgment}
We sincerely thank Dr. M. Emmerich for providing us explicit forms of the
DAs of the $N^\ast$ baryon. One of the authors, T. Barakat, thanks to the
International Scientific Partnership Program ISPP at the King Saud
University for funding his research work through ISPP No: 0038.  
  
\clearpage

\section*{Appendix A: $N^\ast$ distribution amplitudes
}

In this Appendix, we present the $N^\ast$ DAs, which are necessary to
calculate the $\Lambda \to N^\ast$ transition form factors. The DAs of the
$N^\ast$ baryon are defined from the matrix element
$\langle 0 \vel \epsilon^{abc} u_\alpha^a(a_1 x) d_\beta^b(a_2 x)
d_\gamma^c(a_3 x) \ver N^\ast(p)\rangle$. The general decomposition of
this matrix in terms of the DAs of the $N^\ast$ baryon is given below.
(see \cite{Rhly12}),
\bea\label{wave func}
&& 4 \langle 0 \vel \epsilon^{abc} u_\alpha^a(a_1 x) d_\beta^b(a_2 x)
d_\gamma^c(a_3 x) \ver N^\ast(p)\rangle\nnb\\
\es \mathcal{S}_1 m_{N^\ast}C_{\alpha\beta}N_{\gamma}^\ast -
\mathcal{S}_2 m_{N^\ast}^2 C_{\alpha\beta}(\rlap/x N^\ast)_{\gamma}\nnb\\
\ar \mathcal{P}_1 m_{N^\ast}(\gamma_5 C)_{\alpha\beta}(\gamma_5 N^\ast)_{\gamma} +
\mathcal{P}_2 m_{N^\ast}^2 (\gamma_5 C)_{\alpha\beta}(\gamma_5 \rlap/x N^\ast)_{\gamma} -
\left(\mathcal{V}_1 + \frac{x^2 m_{N^\ast}^2}{4} \mathcal{V}_1^M \right) 
(\rlap/p C)_{\alpha\beta} N_{\gamma}^\ast \nnb\\
\ar \mathcal{V}_2 m_{N^\ast}(\rlap/p C)_{\alpha\beta}(\rlap/x N^\ast)_{\gamma} + 
\mathcal{V}_3 m_{N^\ast}(\gamma_\mu C)_{\alpha\beta} (\gamma^\mu N^\ast)_{\gamma} -
\mathcal{V}_4 m_{N^\ast}^2 (\rlap/x C)_{\alpha\beta} N_{\gamma}^\ast \nnb\\
\ek \mathcal{V}_5 m_{N^\ast}^2(\gamma_\mu C)_{\alpha\beta}
(i \sigma^{\mu\nu} x_\nu N^\ast)_\gamma +
\mathcal{V}_6 m_{N^\ast}^3 (\rlap/x C)_{\alpha\beta}(\rlap/x N^\ast)_{\gamma} \nnb \\
\ek \left(\mathcal{A}_1 + \frac{x^2m_{N^\ast}^2}{4}\mathcal{A}_1^M\right)
(\rlap/p \gamma_5 C)_{\alpha\beta} (\gamma N^\ast)_{\gamma} +
\mathcal{A}_2 m_{N^\ast}(\rlap/p \gamma_5 C)_{\alpha\beta} (\rlap/x \gamma_5 N^\ast)_{\gamma} +
\mathcal{A}_3 m_{N^\ast}(\gamma_\mu\gamma_5 C)_{\alpha\beta}
(\gamma^\mu \gamma_5 N^\ast)_{\gamma}\nnb\\
\ek \mathcal{A}_4 m_{N^\ast}^2(\rlap/x \gamma_5 C)_{\alpha\beta}
(\gamma_5 N^\ast)_{\gamma} -
\mathcal{A}_5 m_{N^\ast}^2(\gamma_\mu \gamma_5 C)_{\alpha\beta}
(i \sigma^{\mu\nu} x_\nu \gamma_5 N^\ast)_{\gamma} +
\mathcal{A}_6 m_{N^\ast}^3(\rlap/x \gamma_5 C)_{\alpha\beta}
(\rlap/x \gamma_5 N^\ast)_{\gamma}\nnb\\
\ek \left(\mathcal{T}_1 + \frac{x^2m_{N^\ast}^2}{4}\mathcal{T}_1^M \right)
(i \sigma_{\mu\nu} p_\nu C)_{\alpha\beta} (\gamma^\mu N^\ast)_{\gamma} +
\mathcal{T}_2 m_{N^\ast} (i \sigma_{\mu\nu} x^\mu p^\nu C)_{\alpha\beta}
N_{\gamma}^\ast\nnb\\
\ar \mathcal{T}_3 m_{N^\ast}(\sigma_{\mu\nu} C)_{\alpha\beta}
(\sigma^{\mu\nu} N^\ast)_{\gamma} +
\mathcal{T}_4 m_{N^\ast} (\sigma_{\mu\nu} p^\nu C)_{\alpha\beta}
(\sigma^{\mu\rho} x_\rho N^\ast)_{\gamma} \nnb\\
\ek \mathcal{T}_5 m_{N^\ast}^2 (i\sigma_{\mu\nu} x^\nu C)_{\alpha\beta}
(\gamma^\mu N^\ast)_{\gamma} -  
\mathcal{T}_6 m_{N^\ast}^2 (i \sigma_{\mu\nu} x^\mu p^\nu C)_{\alpha\beta}
(\rlap/x N^\ast)_{\gamma}\nnb\\
\ek \mathcal{T}_7 m_{N^\ast}^2 (\sigma_{\mu\nu} C)_{\alpha\beta}
(\sigma^{\mu\nu} \rlap/x N^\ast)_{\gamma} +
\mathcal{T}_8 m_{N^\ast}^3 (\sigma_{\mu\nu} x^\nu C)_{\alpha\beta}
(\sigma^{\mu\rho} x_\rho N^\ast)_{\gamma}~.\nnb
\eea
The functions labeled with calligraphic letters 
in the above expression do not possess definite twists
but they can be
written in terms of the $N^\ast$ distribution amplitudes (DAs)
with definite and  increasing twists via   the scalar
product $p\mcdot x$ and the parameters $a_i$, $i=1,2,3$. The relations between
the two sets of DAs for the $N^\ast$, and for the scalar, pseudo-scalar, vector, axial
vector and tensor DAs for nucleons are:
\bea
{\cal S}_1 \es S_1 \nnb \\
2 (p\mcdot x) \, {\cal S}_2 \es S_1-S_2 \nnb \\
 {\cal P}_1 \es P_1 \nnb \\
2 (p\mcdot x) \, {\cal P}_2 \es P_2-P_1 \nnb \\
\V_1 \es V_1 \nnb \\
2 (p\mcdot x) \, \V_2 \es V_1 - V_2 - V_3 \nnb \\
2 \V_3 \es V_3 \nnb \\
4 (p\mcdot x) \, \V_4  \es - 2 V_1 + V_3 + V_4  + 2 V_5 \nnb \\
4 (p\mcdot x) \V_5 \es V_4 - V_3 \nnb \\
4 (p\mcdot x)^2  \V_6 \es - V_1 + V_2 +  V_3 +  V_4 + V_5 - V_6 \nnb \\
\A_1 \es A_1 \nnb \\
2 (p\mcdot x) \A_2 \es - A_1 + A_2 -  A_3 \nnb \\
2 \A_3 \es A_3 \nnb \\
4 (p\mcdot x) \A_4 \es - 2 A_1 - A_3 - A_4  + 2 A_5 \nnb \\
4 (p\mcdot x) \A_5 \es A_3 - A_4 \nnb \\
4 (p\mcdot x)^2  \A_6 \es  A_1 - A_2 +  A_3 +  A_4 - A_5 + A_6 \nnb \\
\T_1 \es T_1 \nnb \\
2 (p\mcdot x) \T_2 \es T_1 + T_2 - 2 T_3 \nnb \\
2 \T_3 \es T_7 \nnb \\
 2 (p\mcdot x) \T_4 \es T_1 - T_2 - 2  T_7 \nnb \\
2 (p\mcdot x) \T_5 \es - T_1 + T_5 + 2  T_8 \nnb \\
4 (p\mcdot x)^2 \T_6 \es 2 T_2 - 2 T_3 - 2 T_4 + 2 T_5 + 2 T_7 + 2 T_8 \nnb \\
4 (p\mcdot x) \T_7 \es T_7 - T_8 \nnb \\
4 (p\mcdot x)^2 \T_8 \es -T_1 + T_2 + T_5 - T_6 + 2 T_7 + 2 T_8~ \nnb
\eea
where the terms in $x^2$,  $\mathcal{V}_1^M,\mathcal{A}_1^M$ and $\mathcal{T}_1^M$
are left aside.

The distribution amplitudes $F[a_i(p\mcdot x)]$=  $S_i$, 
$P_i$, $V_i$, $A_i$, $T_i$ can be represented as:
\bea\label{dependent1}
F[a_i (p\mcdot x)]=\int dx_1dx_2dx_3\delta(x_1+x_2+x_3-1) e^{ip\cdot
x\Sigma_ix_ia_i}F(x_i)~.\nnb
\eea
where, $x_{i}$ with $i=1,~2$ and $3$ are longitudinal momentum
fractions carried by the participating quarks.

The explicit expressions for the $\Lambda$ DAs up to twist 6 are given as:\\
Twist--$3$ DAs:
\bea 
\label{twist3}
V_1(x_i,\mu) \es 120 x_1 x_2 x_3 \left[\phi_3^0(\mu) + 
\phi_3^+(\mu) (1- 3 x_3)\right]\,,
\nnb \\
A_1(x_i,\mu) \es 120 x_1 x_2 x_3 (x_2 - x_1) \phi_3^-(\mu) ~,
\nnb \\
T_1(x_i,\mu) \es 120 x_1 x_2 x_3 
\Big[\phi_3^0(\mu) - \frac12\left(\phi_3^+ - \phi_3^-\right)(\mu) 
(1- 3 x_3)\Big]~. \nnb
\eea
Twist--$4$ DAs:
\bea
\label{twist4}
V_2(x_i,\mu)  \es 24 x_1 x_2 
\left[\phi_4^0(\mu)  + \phi_4^+(\mu)  (1- 5 x_3)\right] ~,\nnb\\
A_2(x_i,\mu)  \es 24 x_1 x_2 (x_2 - x_1) \phi_4^-(\mu) ~,\nnb \\
T_2(x_i,\mu) \es 24 x_1 x_2 \left[
\xi_4^0(\mu) + \xi_4^+(\mu)( 1- 5 x_3)\right]~,\nnb \\
V_3(x_i,\mu)  \es  12 x_3 \left[\psi_4^0(\mu)(1-x_3)
+ \psi_4^+(\mu)( 1-x_3 - 10 x_1 x_2)
+ \psi_4^-(\mu) (x_1^2 + x_2^2 - x_3 (1-x_3) ) \right]~,\nnb \\
A_3(x_i,\mu) \es 12 x_3 (x_2-x_1) 
\left[\left(\psi_4^0 + 
\psi_4^+ \right)(\mu)+ \psi_4^-(\mu) (1-2 x_3) \right] ~,\nnb \\
T_3(x_i,\mu)  \es 6 x_3 \left[
(\phi_4^0 + \psi_4^0 + \xi_4^0)(\mu)(1-x_3) +
(\phi_4^+ + \psi_4^+ + \xi_4^+)(\mu) ( 1-x_3 - 10 x_1 x_2)\right. \nnb \\
\ar \left.(\phi_4^- - \psi_4^- + \xi_4^-)(\mu) (x_1^2 + x_2^2 - x_3 (1-x_3) )\right],\nnb \\
T_7(x_i,\mu)  \es 6 x_3 \left[
(\phi_4^0 + \psi_4^0 - \xi_4^0)(\mu)(1-x_3) +
(\phi_4^+ + \psi_4^+ - \xi_4^+)(\mu) ( 1-x_3 - 10 x_1 x_2)\right. \nnb \\
\ar \left.(\phi_4^- - \psi_4^- - \xi_4^-)(\mu) (x_1^2 + x_2^2 - x_3 (1-x_3) )
\right]~,\nnb \\
S_1(x_i,\mu) \es 6 x_3 (x_2-x_1) \left[
(\phi_4^0 + \psi_4^0 + \xi_4^0 + \phi_4^+ + \psi_4^+ 
+ \xi_4^+)(\mu)+ (\phi_4^- - \psi_4^- + \xi_4^-)(\mu)(1-2 x_3) \right]~,\nnb \\
P_1(x_i,\mu) \es 6 x_3 (x_1-x_2) \left[
(\phi_4^0 + \psi_4^0 -\xi_4^0 + \phi_4^+ + \psi_4^+ - 
\xi_4^+)(\mu) + (\phi_4^- - \psi_4^- - \xi_4^-)(\mu)(1-2 x_3) \right]~.\nnb
\eea
Twist--$5$ DAs:
\bea
\label{twist5} 
V_4(x_i,\mu) \es 3 \left[\psi_5^0(\mu)(1-x_3) 
+ \psi_5^+(\mu)(1-x_3 - 2 (x_1^2 +  x_2^2))
+ \psi_5^-(\mu)\left(2 x_1x_2 - x_3(1-x_3)\right) \right]~,\nnb\\
A_4(x_i,\mu) \es 3 (x_2 -x_1)
\left[- \psi_5^0(\mu) + \psi_5^+(\mu)(1- 2 x_3) + \psi_5^-(\mu) x_3\right]~,\nnb \\
T_4(x_i,\mu) \es \frac32 \left[
(\phi_5^0 +  \psi_5^0 + \xi_5^0) (\mu)(1-x_3)
+ \left(\phi_5^+ + \psi_5^+ + \xi_5^+ \right)(\mu)(1-x_3 - 2 (x_1^2 +  x_2^2))
\right. \nnb \\
\ar \left. \left(\phi_5^- - \psi_5^- + \xi_5^- \right) (\mu)\left(2 x_1x_2 -
x_3(1-x_3)\right)\right]~,\nnb \\
T_8(x_i,\mu) \es \frac32 \left[
(\phi_5^0 +  \psi_5^0 - \xi_5^0) (\mu)(1-x_3)
+ \left(\phi_5^+ + \psi_5^+ - \xi_5^+ \right)(\mu)(1-x_3 - 2 (x_1^2 +  x_2^2))
\right. \nnb \\
\ar \left. \left(\phi_5^- - \psi_5^- - \xi_5^- \right) (\mu)\left(2 x_1x_2 - x_3(1-x_3)\right) \right]~, \nnb \\
V_5(x_i,\mu) \es 6 x_3 
\left[\phi_5^0(\mu)  + \phi_5^+(\mu)(1- 2 x_3)\right]~, \nnb\\
A_5(x_i,\mu) \es 6 x_3 (x_2-x_1) \phi_5^-(\mu) ~, \nnb\\
T_5(x_i,\mu) \es 6 x_3 \left[ \xi_5^0(\mu) + \xi_5^+(\mu)( 1- 2 x_3)\right]~,\nnb \\
S_2(x_i,\mu) \es \frac32 (x_2 -x_1) 
\left[- \left(\phi_5^0 + \psi_5^0 + \xi_5^0\right)(\mu)
+ \left(\phi_5^+ + \psi_5^+ + \xi_5^+\right)(\mu) (1- 2 x_3) \right. \nnb \\
\ar \left. \left(\phi_5^- - \psi_5^- + \xi_5^- \right)(\mu) x_3 \right]~, \nnb \\
P_2(x_i,\mu) \es \frac32 (x_2 -x_1)
\left[- \left(-\phi_5^0 - \psi_5^0 + \xi_5^0\right)(\mu)
+ \left(-\phi_5^+ - \psi_5^+ + \xi_5^+\right)(\mu) (1- 2 x_3)\right. \nnb \\
\ar \left. \left(-\phi_5^- + \psi_5^- + \xi_5^- \right)(\mu) x_3  \right]~. \nnb
\eea
Twist-6:
\bea
\label{twist6}
V_6(x_i,\mu) \es 2 \left[\phi_6^0(\mu) +  \phi_6^+(\mu) (1- 3 x_3)\right]~,
\nnb \\
A_6(x_i,\mu) \es 2 (x_2 - x_1) \phi_6^- ~, \nnb \\
T_6(x_i,\mu) \es 2 \Big[\phi_6^0(\mu) - 
\frac12\left(\phi_6^+-\phi_6^-\right) (1\!-\! 3 x_3)\Big]~.\nnb
\eea
Finally the $x^2$ corrections to the corresponding expressions 
${\cal V}_1^M$, ${\cal A}_1^M$, ${\cal T}_1^M$
for the leading twist DAs $V_1$, $A_1$ and $T_1$
in the momentum fraction space are given as:  
\bea
{\cal V}_1^M (x_2) \es \int \limits_0^{1-x_2} dx_1 
V_1^{M}(x_1, x_2,1-x_1-x_2) \nnb \\
\es \frac{x_2^2}{24} \left[ f_{N^\ast} C_{f}^u (x_2) + 
\lambda_1^{N^\ast} C_{\lambda}^u (x_2)\right]~,\nnb
\eea
where
\bea
C_{f}^u (x_2)\es (1 - x_2)^3 \Big[113 + 495x_2 - 552x_2^2 
- 10A_1^u(1 - 3x_2) \nnb \\
\ar  2V_1^d(113 - 951x_2 + 828x_2^2) \Big]~,\nnb \\
C_{\lambda}^u (x_2) \es - (1- x_2)^3 
              \Big[13 - 20f_1^d + 3x_2 + 10f_1^u(1 \!-\! 3x_2)\Big]~. \nnb
\eea
The expression for the axial--vector function ${\cal A}_1^{M(u)} (x_2)$ is given
as:
\bea
{\cal A}_1^{M(u)} (x_2) \es \int_0^{1-x_2} dx_1 A_1^{M}(x_1, x_2,1-x_1-x_2)~,\nnb\\ 
      \es \frac{ x_2^2}{24} (1 - x_2)^3 \left[ f_{N^\ast} D_{f}^u (x_2)  + 
\lambda_1^{N^\ast} D_{\lambda}^u(x_2) \right]~,\nnb 
\eea
with
\bea
D_{f}^u(x_2) \es 11 + 45 x_2 - 2 A_1^u (113 - 951 x_2 + 828 x_2^2 )
+ 10 V_1^d (1 - 30 x_2)~, \nnb \\
D_{\lambda}^u (x_2)& = &  29 - 45 x_2 - 10 f_1^u (7 - 9 x_2) - 
20 f_1^d (5 - 6 x_2)~. \nnb
\eea
Similarly, we get for the function ${\cal T}_1^{M(u)} (x_2)$:
\bea
{\cal T}_1^{M(u)} (x_2) \es \int_0^{1-x_2} dx_1 T_1^{M}(x_1, x_2,1-x_1-x_2)~,\nnb\\ 
\es \frac{ x^2}{48}  \left[ f_{N^\ast} E_{f}^u(x_2)  +
\lambda_1^{N^\ast} E_{\lambda}^u(x_2)  \right]~,\nnb
\eea
where
\begin{eqnarray}
E_{f}^u (x_2) \es -\Big\{
(1 - x_2) \Big[3 (439 + 71 x_2 - 621 x_2^2 + 587 x_2^3 
- 184 x_2^4) \nnb \\
\ar 4 A_1^u (1 - x_2)^2 (59 - 483 x_2 + 414 x_2^2) \nnb \\
\ek 4 V_1^d (1301 - 619 x_2 - 769 x_2^2 + 1161 x_2^3 - 414 x_2^4)\Big]
\Big\}
- 12 (73 - 220V_1^d) \ln[x_2]~, \nnb \\
E_{\lambda}^u (x_2) \es -\Big\{(1 - x_2) 
\Big[5 - 211 x_2 + 281 x_2^2 - 111 x_2^3 
+ 10 (1 + 61 x_2 - 83 x_2^2 + 33 x_2^3) f_1^d \nnb \\
\ek 40(1 - x_2)^2 (2 - 3 x_2) f_1^u\Big]
\Big\} - 12 (3 - 10 f_1^d) \ln[x_2]~.\nnb
\eea

The following functions are encountered to the above amplitudes and they
can be  defined in
terms of the  8 independent parameters, namely  $f_{N^\ast}$,
$\lambda_1$, $\lambda_2$ and $f_1^u,f_1^d,f_2^d,A_1^u,V_1^d$:\\

\bea
\label{DAs}
\phi_3^0 \es \phi_6^0 = f_{N^\ast} \nnb \\
\phi_4^0 \es \phi_5^0 = {1 \over 2} (f_{N^\ast} + \lambda_1^{N^\ast}) \nnb \\
\xi_4^0 \es \xi_5^0 = {1 \over 6} \lambda_2^{N^\ast} \nnb \\
\psi_4^0 \es \psi_5^0 = {1 \over 2} (f_{N^\ast} - \lambda_1^{N^\ast})~, \nnb\\
\phi_3^- \es {21 \over 2} f_{N^\ast} A_1^u,~~\phi_3^+ = {7 \over 2}
f_{N^\ast} (1 - V_1^d)~, \nnb \\
\phi_4^+ \es {1\over 4} \left[f_{N^\ast} (3-10 V_1^d) + \lambda_1^{N^\ast}
(3 - 10 f_1^d) \right]~, \nnb \\
\phi_4^- \es - {5\over 4} \left[f_{N^\ast} (1-2 A_1^u) - \lambda_1^{N^\ast}
(1 - 2 f_1^d - 4 f_1^u) \right]~, \nnb \\
\psi_4^+ \es - {1\over 4} \left[f_{N^\ast} (2 + 5 A_1^u - 5 V_1^d) - 
\lambda_1^{N^\ast} (2 - 5 f_1^d - 5 f_1^u) \right]~, \nnb \\
\psi_4^- \es {5\over 4} \left[f_{N^\ast} (2 - A_1^u - 3 V_1^d) - 
\lambda_1^{N^\ast} (2 - 7 f_1^d + f_1^u) \right]~, \nnb \\
\xi_4^+ \es {1\over 16} \lambda_2^{N^\ast} (4 - 15 f_2^d)~, \nnb \\
\xi_4^- \es {5\over 16} \lambda_2^{N^\ast} (4 - 15 f_2^d)~, \nnb \\
\phi_5^+ \es  - {5\over 6} \left[f_{N^\ast} (3 + 4 V_1^d) - \lambda_1^{N^\ast}
(1 - 4 f_1^d)\right]~, \nnb \\
\phi_5^- \es - {5\over 3} \left[f_{N^\ast} (1 - 2 A_1^u) -
\lambda_1^{N^\ast} (f_1^d - f_1^u)\right]~, \nnb \\
\psi_5^+ \es - {5\over 6} \left[f_{N^\ast} (5 + 2 A_1^u - 2 V_1^d) - 
\lambda_1^{N^\ast} (1 - 2 f_1^d - 2 f_1^u) \right]~,\nnb \\
\psi_5^- \es {5\over 3} \left[f_{N^\ast} (2 - A_1^u - 3 V_1^d) +     
\lambda_1^{N^\ast} (f_1^d - f_1^u) \right]~, \nnb \\
\xi_5^+ \es {5\over 36} \lambda_2^{N^\ast}  (2 - 9 f_2^d)~,\nnb \\
\xi_5^- \es -{5\over 4} \lambda_2^{N^\ast} f_2^d~, \nnb \\
\phi_6^+ \es  {1\over 2} \left[f_{N^\ast} (1 - 4 V_1^d) -
\lambda_1^{N^\ast} (1 - 2 f_1^d)\right]~, \nnb \\
\phi_6^- \es {1\over 2} \left[f_{N^\ast} (1 + 4 A_1^d) +                   
\lambda_1^{N^\ast} (1 - 4 f_1^d - 2 f_1^u)\right]~, \nnb
\eea
where the parameters $A_1^u,~V_1^d,~f_1^d,~f_1^u$ and $f_2^d$ are defined as
\cite{Rhly16},
\bea
A_1^u \es \varphi_{10} + \varphi_{11}~,\nnb \\
V_1^d \es {1\over 3} - \varphi_{10} + {1\over 3} \varphi_{11}~,\nnb \\
f_1^u \es {1\over 10} - {1\over 6} {f_{N^\ast} \over \lambda_1^{N^\ast}} -
{3\over 5} \eta_{10} - {1\over 3} \eta_{11}~, \nnb \\
f_1^d \es {3\over 10} - {1\over 6} {f_{N^\ast} \over \lambda_1^{N^\ast}} +  
{1\over 5} \eta_{10} - {1\over 3} \eta_{11}~, \nnb \\
f_2^d \es {4\over 15} + {2\over 5} \xi_{10}~,\nnb
\eea
The numerical values of the parameters 
$\varphi_{10},~\varphi_{11},~\varphi_{20},~\varphi_{21},~\varphi_{22},
~\eta_{10},~\eta_{11}$ and $f_{N^\ast}/\lambda_1^{N^\ast}$,
and $\lambda_1^{N^\ast}/\lambda_1^N$ are presented in Table 4 
(this Table is taken from \cite{Rhly10}).

\begin{table}[h]

\renewcommand{\arraystretch}{1.3}
\addtolength{\arraycolsep}{-0.5pt}
\small
$$
\begin{array}{|l|c|c|c|c|c|c|c|c|c|c|}
\hline \hline
& \mbox{Model} & \lambda_1^{N^\ast}/\lambda_1^N &
f_{N^\ast}/\lambda_1^{N^\ast} & \varphi_{10} & \varphi_{11} & \varphi_{20} &
\varphi_{21} & \varphi_{22} & \eta_{10} & \eta_{11} \\  \hline
\cite{Rhly16} &\mbox{ LCSR--1} & 0.633 & 0.027 & 0.36 & -0.95 & 0 & 0 & 0 & 0     & 0.94 \\
\cite{Rhly16} &\mbox{LCSR--2}  & 0.633 & 0.027 & 0.37 & -0.96 & 0 & 0 & 0 & -0.29 & 0.23 \\
\cite{Rhly17} &\mbox{LATTICE} & 0.633 & 0.027 & 0.28 & -0.86 & 1.7 & -2 & 1.7 &  0 & 0 \\
\hline \hline
\end{array}
$$
\caption{Parameters of the DAs for the $N^\ast(1535)$ baryon at $\mu^2 =
2~GeV^2$}
\renewcommand{\arraystretch}{1}
\addtolength{\arraycolsep}{-1.0pt}
\end{table}

\clearpage


\section*{Appendix B: Differential widths for the
$\Lambda_{b}\to N^\ast \ell^{+}\ell^{-}$ decays}
\setcounter{equation}{0}
\setcounter{section}{0}


In this Appendix we present the differential widths for the
$\Lambda_{b}\to N^\ast \ell^{+}\ell^{-},~(\ell=e,\mu,\tau)$ decays.
After lengthy, but straightforward
calculations, for the differential rate of the $\Lambda_{b}\to
N^\ast \ell^{+}\ell^{-}$ we get,

\bea
\frac{d\Gamma(s)}{ds} = \frac{G^2\alpha^2_{em} m_{\Lambda_b}}{4096 \pi^5}
\vel V_{tb}V_{td}^\ast\ver^2 v \sqrt{\lambda(1,r,s)} \, \Bigg[  T_1(s) +
\frac{1}{3} T_2(s)\Bigg]~, \nnb
\eea
where $s= q^2/m^2_{\Lambda_b}$, $r= m_{N^\ast}^2/m^2_{\Lambda_b}$,
$G_F = 1.17 \times 10^{-5}$ GeV$^{-2}$ is the Fermi
coupling constant, $v=\sqrt{1-4 m_\ell^2/q^2}$ is the
lepton velocity, and $\lambda(a,b,c)=a^2+b^2+c^2-2ab-2ac-2bc$ is the usual
triangle function. For the element of the CKM matrix
$\vel V_{tb}V_{td}^\ast \ver = (8.2 \pm 0.6)\times 10^{-3}$ has been used
\cite{Rhly13}. The functions $T_1(s)$ and $T_2(s)$ are given as:

\bea
T_1(s) \es
%
8  m_{\Lambda_b}^2 \Bigg\{
(1 + 2 \sqrt{r} + r - s) \left[4 m_\ell^2  + m_{\Lambda_b}^2  (1 - 2 \sqrt{r} + r + s)\right]
\ve F_1 \ve^2 \nnb \\
\ek \Big[4 m_\ell^2  (1 + 6 \sqrt{r} + r - s) -  
        m_{\Lambda_b}^2  \Big( (1 - r)^2  + 4 \sqrt{r} s - s^2 \Big)\Big]
 \ve F_4 \ve^2 \nnb \\
\ar (1 + 2 \sqrt{r} + r - s) 
    \Big[4 m_\ell^2  (1 - \sqrt{r})^2  + m_{\Lambda_b}^2  s (1 - 2 \sqrt{r} + r +
     s)\Big] \ve F_2 \ve^2 \nnb \\
\ar m_{\Lambda_b}^2  s \Big[(1 - r)^2  + 4 \sqrt{r} s - s^2 \Big] v^2
  \ve F_4 \ve^2 \nnb \\
\ar 4 m_\ell^2  (1 - 2 \sqrt{r} + r - s) s \ve F_6 \ve^2 \nnb \\
\ar (1 - 2 \sqrt{r} + r - s) \Big[4 m_\ell^2  + m_{\Lambda_b}^2  (1 + 2 \sqrt{r} + r + s) \Big] 
    \ve G_1 \ve^2 \nnb \\
\ek \Big[ 4 m_\ell^2  (1 - 6 \sqrt{r} + r - s) - 
        m_{\Lambda_b}^2  \Big( (1 - r)^2  - 4 \sqrt{r} s - s^2 \Big)\Big]
 \ve G_4 \ve^2 \nnb \\
\ar (1 - 2 \sqrt{r} + r - s) 
    \Big[4 m_\ell^2  (1 + \sqrt{r})^2  + m_{\Lambda_b}^2  s (1 + 2 \sqrt{r} + r +
s)\Big] \ve G_2 \ve^2\nnb \\
\ar m_{\Lambda_b}^2  s \Big[(1 - r)^2  - 4 \sqrt{r} s - s^2 \Big] v^2  \ve
G_5 \ve^2 \nnb \\
\ar 4 m_\ell^2  (1 + 2 \sqrt{r} + r - s) s \ve G_6 \ve^2 \nnb \\
\ek 4 (1 - \sqrt{r}) (1 + 2 \sqrt{r} + r - s) (2 m_\ell^2  + m_{\Lambda_b}^2  s) 
\mbox{\rm Re}[F_1^\ast F_2] \nnb \\
\ek 4 m_{\Lambda_b}^2  (1 - \sqrt{r}) (1 + 2 \sqrt{r} + r - s) s v^2  
\mbox{\rm Re}[F_4^\ast F_5] \nnb\\
\ek 8 m_\ell^2  (1 + \sqrt{r}) (1 - 2 \sqrt{r} + r - s) 
 \mbox{\rm Re}[F_4^\ast F_6] \nnb\\   
\ek 4 (1 + \sqrt{r}) (1 - 2 \sqrt{r} + r - s) (2 m_\ell^2  +
m_{\Lambda_b}^2  s) \mbox{\rm Re}[G_1^\ast G_2] \nnb\\   
\ek 4 m_{\Lambda_b}^2  (1 + \sqrt{r}) (1 - 2 \sqrt{r} + r - s) s v^2  
\mbox{\rm Re}[G_4^\ast G_5] ] \nnb\\   
\ek 8 m_\ell^2  (1 - \sqrt{r}) (1 + 2 \sqrt{r} + r - s) 
\mbox{\rm Re}[G_4^\ast G_6]
\Bigg\}~, \nnb
\eea
\bea
T_2 (s) \es
- 8 m_{\Lambda_b}^4  v^2 \lambda(1,r,s) \Big[
\ve F_1 \ve^2  + \ve F_4 \ve^2  + 
\ve G_1 \ve^2  + \ve G_4 \ve^2  - 
s \Big(\ve F_2 \ve^2 + \ve F_5 \ve^2  + 
\ve G_2 \ve^2  + \ve G_5 \ve^2 \Big)\Big]~. \nnb
\eea
%
%
The differential decay width for the $\Lambda_b^\ast \rar N^\ast \ell^+\ell^-$
transition cam be obtained from the differential decay width for the
$\Lambda_b \rar N^\ast \ell^+\ell^-$ by making the following replacements:
$F_i \rar \widetilde{G}_i$, $G_i \rar \widetilde{F}_i$, $ m_N \rar - m_N$,
and by changing the sign in front of the terms $\mbox{\rm Re}[F_4^\ast
F_5]$, $\mbox{\rm Re}[F_4^\ast F_6]$, and $\mbox{\rm Re}[G_4^\ast G_5]$,
as well as $m_{\Lambda_b} \rar m_{\Lambda_b^\ast}$, $s \rar
s^\prime = q^2/m_{\Lambda_b^\ast}^2$, and $r \rar r^\prime =
m_{N^\ast}^2/m_{\Lambda_b^\ast}^2$. 
%
%
where
\bea
F_1(q^2) \es \phantom{-} c_9 g_1(q^2) - {2 m_b\over m_{\Lambda_b^\ast}} c_7 g_1^T(q^2)~, \nnb \\
F_2(q^2) \es - c_9 g_2(q^2) - {2 m_b\over q^2} m_{\Lambda_b^\ast} c_7 g_2^T(q^2)~, \nnb \\
F_3(q^2) \es - c_9 g_3(q^2) - {m_b\over q^2} (m_{\Lambda_b^\ast}-m_N) c_7 g_1^T(q^2)~, \nnb \\
G_1(q^2) \es \phantom{-} c_9 f_1(q^2) - {2 m_b\over m_{\Lambda_b^\ast}} c_7 f_1^T(q^2) ~, \nnb \\
G_2(q^2) \es - c_9 f_2(q^2) - {2 m_b\over q^2} m_{\Lambda_b^\ast} c_7 f_2^T(q^2)~, \nnb \\
G_3(q^2) \es - c_9 f_3(q^2) - {2 m_b\over q^2} (m_{\Lambda_b^\ast}+m_N) c_7 f_1^T(q^2)~, \nnb \\
F_4(q^2) \es \phantom{-} c_{10} g_1(q^2)~, \nnb \\
F_5(q^2) \es - c_{10} g_2(q^2)~, \nnb \\
F_6(q^2) \es - c_{10} g_3(q^2)~, \nnb \\
G_4(q^2) \es \phantom{-} c_{10} f_1(q^2)~, \nnb \\
G_5(q^2) \es - c_{10} f_2(q^2)~, \nnb \\
G_6(q^2) \es - c_{10} f_3(q^2)~. \nnb
\eea

\newpage

\section*{Figure captions}
{\bf Fig. (1)} The dependence of the differential branching ratio 
for the $\Lambda_b \rar N^\ast \mu^+ \mu^-$ transition on $s$,
at $s_0=40~GeV^2$, and $M^2=25~GeV^2$. \\ \\
{\bf Fig. (2)} The same as in Fig. (1), but for the $\Lambda_b \rar N^\ast \tau^+
\tau^-$ transition. \\ \\

\newpage

\begin{figure}[t]
\vskip -1.5cm
\begin{center}
\scalebox{0.785}{\includegraphics{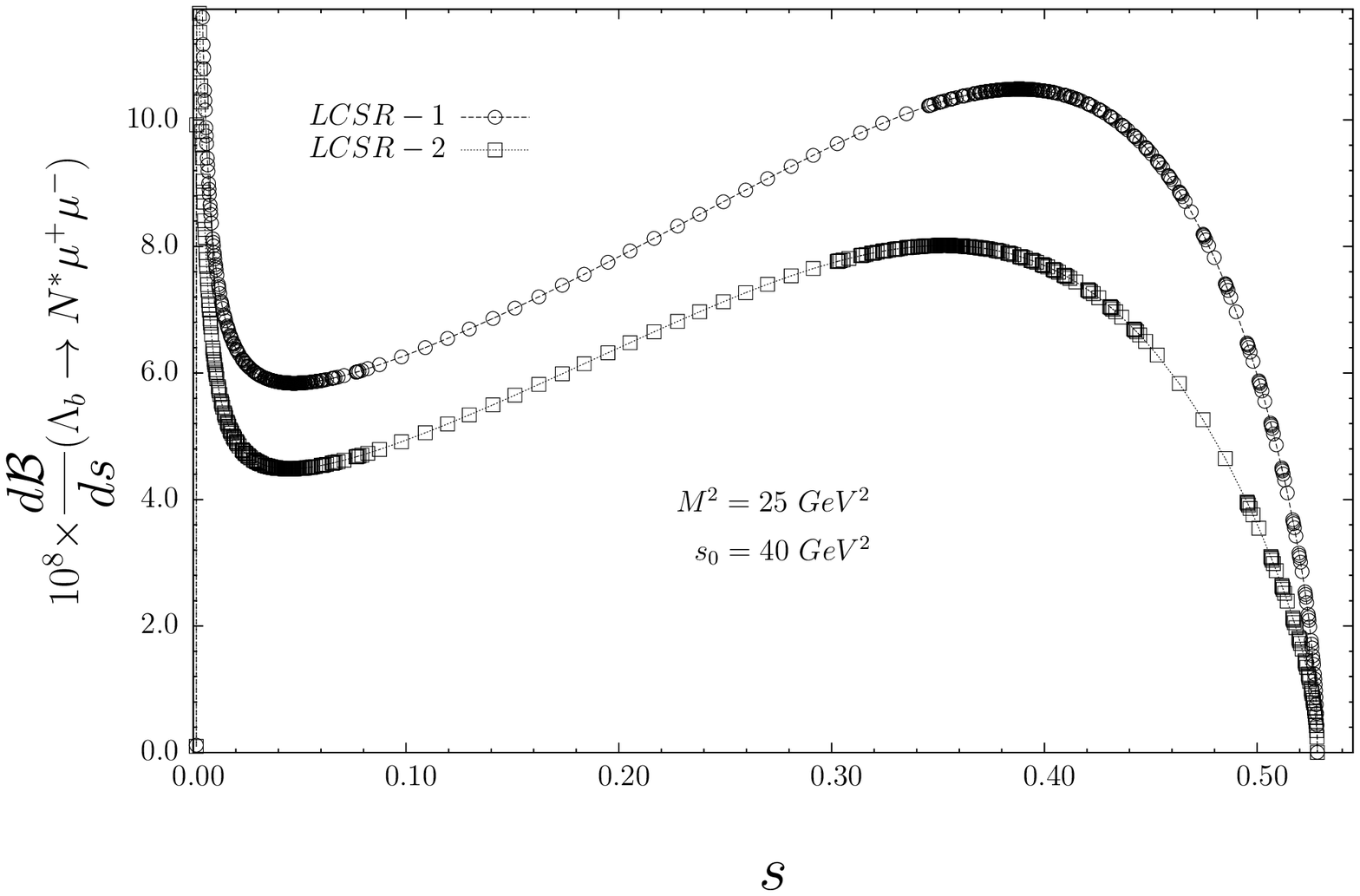}}
\end{center}
\vskip -12.0cm
\caption{}
\end{figure}

\begin{figure}[b]
\vskip -1.5cm
\begin{center}
\scalebox{0.785}{\includegraphics{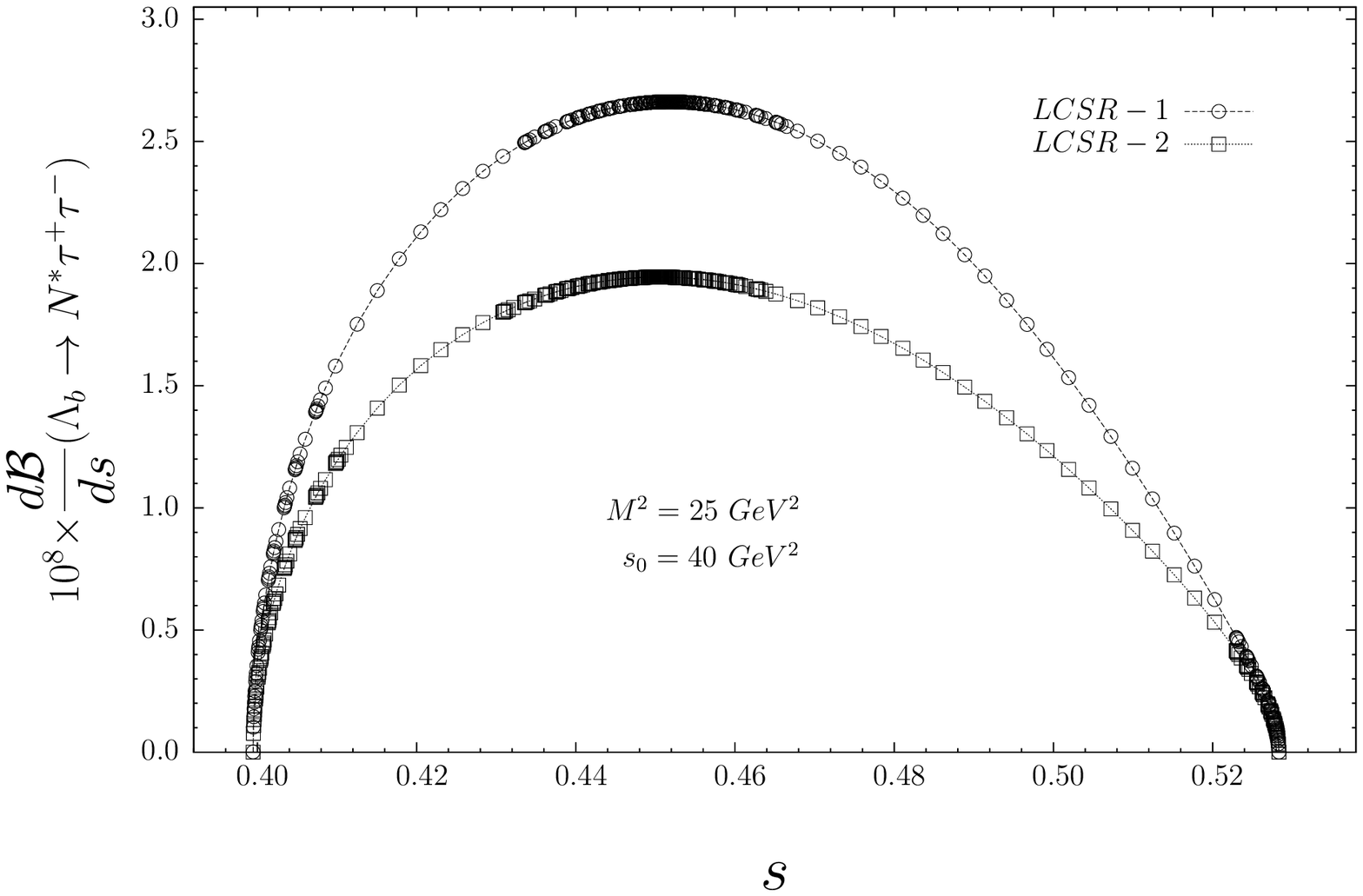}}
\end{center}
\vskip -12.0cm
\caption{}
\end{figure}

\end{document}